\documentclass[pra,twocolumn,superscriptaddress,showpacs]{revtex4}
\usepackage{amsmath,float}
\usepackage{graphicx}
\renewcommand{\eqref}[1]{Eq.~(\ref{#1})}
\newcommand{\half}{\ensuremath{\tfrac{1}{2}}}
\newcommand{\dd}{\,\text{d}}
\newcommand{\nn}{\nonumber}
\newcommand{\ih}{\frac{1}{i\hbar}}

\newcommand{\der}{\frac{d}{dt}}
\newcommand{\INF}{\int_{-\pi}^{\pi}}
\newcommand{\THETA}{\frac{\theta}{2}}

\newcommand{\N}[1][]{\ensuremath{|n #1\rangle \,}}
\newcommand{\NN}[1][]{\ensuremath{\langle n #1}}
\newcommand{\EX}[1][]{\ensuremath{|e #1\rangle \,}}
\newcommand{\GR}[1][]{\ensuremath{|g #1\rangle \,}}
\newcommand{\PLUS}[1][]{\ensuremath{|+ #1\rangle \,}}
\newcommand{\MINUS}[1][]{\ensuremath{|- #1\rangle \,}}
\newcommand{\PLUSM}[1][]{\ensuremath{| \pm #1\rangle \,}}

\begin{document}
        \title{Geometric phase for an adiabatically evolving 
        open quantum system} \author{Ingo Kamleitner}
        \affiliation{Australian Centre for Quantum Computer Technology,
        Macquarie University, Sydney, New South Wales 2109, Australia}
        \author{James~D.~Cresser} \affiliation{Department of Physics,
        Macquarie University, Sydney, New South Wales 2109, Australia.}
        \author{Barry~C.~Sanders} \affiliation{Australian Centre for
        Quantum Computer Technology, Macquarie University, Sydney, New
        South Wales 2109, Australia} \affiliation{Institute for Quantum
        Information Science, University of Calgary, Alberta T2N 1N4,
        Canada} \date{\today}
    \begin{abstract}
      We derive an elegant solution for a two-level system evolving
      adiabatically under the influence of a driving field with a
      time-dependent phase, which includes open system effects such
      as dephasing and spontaneous emission.  This solution, which is
      obtained by working in the representation corresponding to the
      eigenstates of the time-dependent Hermitian Hamiltonian,
      enables the dynamic and geometric phases of the evolving
      density matrix to be separated and relatively easily calculated.
        \end{abstract}
\maketitle

\section{Introduction}

The discovery by Berry \cite{Berry, Simon} that a (non-degenerate)
state of a quantum system can acquire a phase of purely geometric
origin when the Hamiltonian of the system undergoes a cyclic, adiabatic
change has lead to an explosion of interest in this and related
geometric phases in quantum mechanics, both from a theoretical
perspective, and from the point of view of possible applications, the
latter including applications to optics (where the geometric phase was
first discovered~\cite{Pan56}), NMR and molecular physics, and to
quantum computing \cite{JonesVAG,EkertEHIJOV}.  Since Berry's work, and
the demonstration that Berry's phase can be understood as a holonomy
associated with the parallel transport of the quantum state
\cite{Simon}, there have been numerous proposals for generalizations.
The first of these was due to Wilczek and Zee \cite{WilczekZee} who, by
considering a Hamiltonian with non-degenerate eigenstates, established
the existence of an intimate connection between Berry's phase and
non-Abelian gauge theories.  The restriction to changes occurring
adiabatically was relaxed in the work of Aharonov and Anandan
\cite{Aha87} while Anandan \cite{Anandan} generalized the
geometric phase to the non-adiabatic non-Abelian case.  The restriction
of cyclicity was removed by Samuel and Bhandari \cite{SamuelBhandari}
and by Pati \cite{Pati}.  All of this work is concerned with geometric
phases of pure states of closed systems and is now standard, though
\cite{SamuelBhandari} indicated extensions to taking account of quantum
measurements and consequent non-unitary evolution. A nice 
overview, theoretical as well as experimental, is given in 
\cite{Ben-Aryeh}. 

More recently attention has turned to studying geometric phases for 
mixed states, though there is not yet a standard description for 
geometric phases associated with mixed states.

As realistic systems always interact with their environment, and as an
open system is almost always to be found in a mixed state, open systems
are a natural source of problems involving the geometric phases of
mixed states.  Garrison and Wright \cite{GarissonWright} were the first
to touch on this issue in a phenomenological way, by describing open
system evolution in terms of a non-Hermitean Hamiltonian.  This was, in
fact, a pure state analysis, so it did not, strictly speaking, directly
address the problem of geometric phases for a mixed state, but this
work raised issues which could potentially have a bearing on the
analysis of the mixed state problem.  In fact, they did point out that
a proper treatment of an open system would require making use of the
density operator approach.  Nevertheless, they arrived at an
interesting result, a \emph{complex} geometric phase for dissipative
evolution.  This is a result that has been recently put into doubt by a
master equation treatment by Fonseca Romero et al and Aguiar Pinto and
Thomaz \cite{Fonseca,Pin03}.

The first complete open systems analyses of geometric phase for a mixed
state, from two different perspectives, is to be found in the papers of
Ellinas et al \cite{Ell89} and Gamliel and Freed \cite{Gam89}.  The
former worked with the standard master equation for the density
operator of a multilevel atom subject to radiative damping and driven
by a laser field with a time-dependent phase.  What is of interest in
their approach is that it entailed introducing eigenmatrices of the
Liouville superoperator of the master equation for the damped system.
The system Hamiltonian was allowed to vary adiabatically, with the
result that a non-degenerate eigenmatrix acquires a geometric phase as
well as a dynamic phase.  In \cite{Gam89}, the effect of the
environment was modelled as an external classical stochastic influence
which, when averaged, gives rise to the relaxation terms of the
master equation for the system.  In both cases the effects of any
geometric phase was then shown to be present in measurable quantities
such as the inversion of a two state system.

Since then, research has been increasing rapidly into the problem of
defining a geometric phase for mixed quantum states for both unitary
and non-unitary evolution, motivated to a very large extent by the need
to understand the effects of decoherence in quantum computational
processes that exploit geometric phases as a means of constructing
intrinsically fault-tolerant quantum logic gates.  This issue has been
addressed from two points of view, the first holistic in nature wherein
the aim is to identify a geometric phase to be associated with the
mixed state itself, and the second, essentially the approach of
\cite{Ell89}, which works with the pure state geometric phases of an
appropriate set of parallel transported basis states, which then gives
rise to geometric phase factors in the off-diagonal elements of the
density operator.  No geometric phase is explicitly associated with the
mixed state itself; instead observable quantities that will exhibit the
effects of the geometric phases of the underlying basis states are
determined.

The former, holistic approach was first introduced in a formal way by
Uhlmann \cite{Uhlmann}, and in a different way, based on
phase-sensitive measurements via interferometry, by Sj\"{o}qvist et al
\cite{SPEAEOV} for unitary evolution of a mixed state, and later for
non-unitary evolution \cite{ESBOP,deFaria}.  The phase defined in this
way is not the same in all respects to that proposed by Uhlmann
\cite{Tidstrom}.

The latter kind of approach has been used only for open systems, and
involves working with the (Markovian) master equation of the open
system.  The approaches used involve either solving the master
equation of the system \cite{Ell89,Gam89,Fonseca,Pin03}, or
employing a quantum trajectory analysis \cite{Nazir,Carollo,Mar04} to
unravel the dynamics into pure state trajectories, and calculating the
geometric phases associated with individual pure state trajectories.
Noise of a classical origin, such as stochastic fluctuations of the
parameters of the Hamiltonian have also been studied by Chiara and
Palma \cite{ChiaraPalma}.  In essence, the common feature is not so
much to propose a new definition of geometric phase for a mixed state
as to show how the underlying existence of a geometric phase will
nevertheless show up in the observed behaviour of an open quantum
system.  It is this perspective that is adopted in the work to be 
presented in this paper.

Here we introduce an elegant approach for solving the central master
equation which is based on introducing a unitary transformation due to
Kato \cite{Kato} described in the classic text by Messiah
\cite{Messiah}.  

The new picture is defined via a time-dependent unitary transformation
$A^\dagger(t)$, usually referred to as a rotating axis transformation,
which is such that the transformed system Hamiltonian has time
independent eigenspaces.  This method is extended by showing that
under the conditions of adiabatic evolution, all the information on
geometric phase for a closed loop is contained within $A(t)$, and is
regained by transforming back to the original picture.  The goals of
this approach is its simplicity, since one needs only to calculate the
geometric phases for the eigenstates of the Hamiltonian $H(t)$, and
the fact that dynamic and geometric phase are separated in a clear
way.  In fact, perfect separation of geometric and dynamical
contributions is obtained provided the Hamiltonian evolution is
adiabatic and the coupling to the environment is weak.  This approach
bears some similarity to that used by Fonseca RomeroFonseca\cite{Fonseca}
who make use of several unitary transformations to separate the
geometric phase from the dynamic phase.  However, the rationale for their
transformations, and the origin of the geometric phase, is somewhat elusive
in their analysis.  In contrast to \cite{Fonseca}, with the transformation
introduced here, the parallel transport condition is essential and
explains the appearance of the geometric phase.

Within the approach used here, it is possible to show explictly how to
achieve, under certain circumstances, operational cancellation of the
dynamic phase, thereby making the geometric phase accessible in
experiments.  An example of where this is possible is given in
Section~\ref{Examples}.

This paper is organized as follows.  In Section~\ref{MathMethods} we
present the main ideas.  In Section~\ref{Examples} we look at several
examples.  In Section~\ref{conclusions} we summarize our results while
in Section~\ref{newdirections} we indicate possible new directions,
including generalisations to non-cyclic evolution and non-Abelian
holonomies.  An analog to the adiabatic theorem is proved for a general
Lindblad equation \cite{Lin76} in the Appendix.

\section{The Rotating Axis Transformation for Nondegenerate Multilevel
Systems
\label{MathMethods}}

\subsection{Geometric Phases for a Closed System}

For the present we consider the case of a closed system so as to
introduce the basic method employed here.  Suppose we have a system
with Hamiltonian $H(t)$, a function of time due to the dependence of
$H$ on parameters whose values can be changed in time.  This
Hamiltonian will have instantaneous eigenvectors $\N[(t)]$ with
eigenvalues $E_{n}(t)$:
\begin{equation} 
    H(t) \N[(t)] = E_n(t) \N[(t)].
\end{equation}
For simplicity we assume $E_i(t) \neq E_j(t)$ for $i\neq j$.  This
restriction will be removed in section IV to obtain non-Abelian
holonomies.  For an adiabatically slow change in the system 
parameters,
these eigenvectors will also change in such as way as to satisfy the
parallel transport condition \cite{Simon}:
\begin{equation}
     \NN[(t)] | \der\N[(t)]=0.
     \label{parallel} 
\end{equation} 
At this point we introduce a unitary operator $A(t)$ via the equation
\begin{equation} 
    A(t)\N[(0)] = \N[(t)].
    \label{defnofA}
\end{equation}
This completely defines $A(t)$.  Note that because of the path
dependence of the parallel transported eigenstates \N[(t)], the
operator $A(t)$ has a non-integrable nature, and, as we see later, will
contain the information on the geometric phase.

This unitary operator can now be used to remove the time dependence of
the eigenstates of the Hamiltonian.  Thus, if we define
\begin{equation}
    H^{A} \equiv A^{\dagger}HA
\end{equation}
we note that the eigenvectors of $H^{A}$ are now just $\N[(0)]$ and
hence are time independent.  The transformed Schr\"{o}dinger equation
is then
\begin{equation}    
        H^{A}|\psi^{A}\rangle=i\hbar\left(A^{\dagger}\dot{A}|\psi^{A}\rangle
    +\frac{d}{dt}|\psi^{A}\rangle\right)
\end{equation}
where $|\psi^{A}\rangle=A^{\dagger}|\psi\rangle$.  If \N[(t)] is
parallel transported, then, to the lowest-order
adiabatic approximation, one can neglect the terms containing
$A^\dagger (t) \dot{A}(t)$ so that the transformed
Schr\"{o}dinger equation becomes~\cite{Messiah}
\begin{equation}
    H^{A}|\psi^{A}\rangle=i\hbar\frac{d}{dt}|\psi^{A}\rangle.
    \label{purestatenogeom}
\end{equation} 
The solution of \eqref{purestatenogeom} contains no geometric
contribution --- it gives the dynamic contribution to the phase of
any adiabatically evolving state.  So we have extracted the dynamics
from the geometric influence of the time-varying Hamiltonian.  The
geometric contribution is entirely contained within the operator $A$.
To obtain this information we have to transform back to the original
picture in terms of the states $|\psi\rangle$:
\begin{equation}
    |\psi\rangle=A|\psi^{A}\rangle.
\end{equation}
If the Hamiltonian undergoes a closed loop in time $T$, i.e.
$H(T)=H(0)$, then the parallel transported eigenstates \N[(T)] return
to the initial eigenstates \N[(0)] up to the geometric phase.  Hence we
have $A(T)= \text{diag} (e^{i\varphi_1}, \ldots ,e^{i\varphi_N})$ where
$\varphi_n$ is the geometric phase associated with the eigenstate
\N[(T)].  This result can be readily generalized if the system is in a
mixed state $\rho$.  Introducing the notation
\begin{equation}
        \rho^A(t) = A^\dagger(t)\rho(t)A(t)
\end{equation}
we obtain
\begin{align} 
        \rho(T) =& A(T)\rho^A(T)A^\dagger(T) \nonumber \\ 
        =& \text{diag}  
(e^{i\varphi_1},\ldots,e^{i\varphi_N})\rho^A(T)\nonumber\\ 
        &\times \text{diag}(e^{-i\varphi_1}, \ldots ,e^{-i\varphi_N}).
        \label{rhoholonomy}
\end{align}
This holonomic transformation multiplies the off-diagonal elements of
the density operator $\rho^A_{ij}$ by a phase
$e^{i(\varphi_j-\varphi_i)}$, which is the difference of the geometric
phases of the eigenstates of the Hamiltonian $H(T)$.

\subsection{Open System and Master Equation}

Systems that are coupled to a reservoir (or environment) can usually be
described by a reduced density operator that evolves according to a
master equation, which, in many cases, can be written in the Lindblad
form \cite{Lin76}
\begin{equation} 
    \dot{\rho} (t) = \ih [H(t),\rho (t)] +\frac{1}{2}
    \sum_{\alpha=1}^k \mathcal{L}_{\Gamma_\alpha} [\rho(t)]
    \label{meqn1}
\end{equation}
for
\begin{equation}
        \mathcal{L}_\Gamma[\rho] \equiv 2 \Gamma \rho \Gamma^\dagger
        -\Gamma^\dagger \Gamma \rho - \rho\Gamma^\dagger \Gamma~,
\end{equation}
and where $\rho(t)$ is the density operator for the system of interest,
$\dot{\rho}(t)$ is its derivative with respect to time, $H(t)$ is the
system Hamiltonian and the dissipation operators $\Gamma_\alpha$
arise due to the presence of the reservoir.  To obtain a geometric phase
we once again consider an adiabatically changing Hamiltonian $H(t)$,
with $H(T)=H(0)$, as in the preceding section.  Upon introducing the
operator $A(t)$ defined in \eqref{defnofA} we can transform the master
equation with the unitary operator $A^\dagger(t)$, which leads us to
the new master equation
\begin{align} 
    \dot{\rho}^A (t) = &\ih [H^A(t),\rho^A (t)]
    +\rho^A(t)A^\dagger (t) \dot{A}(t)\nonumber \\ & -A^\dagger (t)
    \dot{A}(t)\rho^A(t) +\half\sum_{\alpha =1}^k
    \mathcal{L}_{\Gamma_\alpha^A} [\rho^A(t)] . \label{meqn2}
\end{align}
Note that $\Gamma_\alpha^A\equiv A^\dagger\Gamma_\alpha A$ is in
general time-dependent even when $\Gamma_\alpha$ is not.  As shown in
the Appendix, the terms containing $A^\dagger (t) \dot{A}(t)$ can be
neglected, as in the case of a unitary evolution but with the
additional requirement that the damping is weak.  This establishes the
parallel transport condition \eqref{parallel} for weakly damped systems
in an adiabatic evolution.  We then get
\begin{equation} 
    \dot{\rho}^A (t) = \ih [H^A(t),\rho^A (t)] +
    \half\sum_{\alpha =1}^k  \mathcal{L}_{\Gamma_\alpha^A} 
[\rho^A(t)] .
    \label{meqn3}
\end{equation}
%Furthermore, for two level systems and under some restrictions for
%many level systems, we show in the Appendix that in the case of a
%diagonal Hamiltonian $H^A(t)$, only the absolute value of the
%off-diagonal elements of $ \Gamma^A_\alpha $ are essential.  Since the
%eigenspaces of $H^A(t)$ are time independent we can always achieve a
%diagonal Hamiltonian by means of a proper time independent
%transformation.  This considerably simplifies solving the master
%equation as will be seen in the examples and as discussed in more
%detail in the Appendix.

The solution of \eqref{meqn3} contains no geometric
contributions.  To regain $\rho(T)$ we have to transform back to the
original picture as in the unitary case, \eqref{rhoholonomy}.
\begin{align} 
        \rho(T) =& A(T)\rho^A(T)A^\dagger(T) \nonumber \\ 
    =& \text{diag}
    (e^{i\varphi_1}, \ldots ,e^{i\varphi_N})\rho^A(T)\nonumber \\
        & \times \text{diag}(e^{-i\varphi_1}, \ldots ,e^{-i\varphi_N})
\end{align}
with $\varphi_n$ being the geometric phase for the state \N .

\section{Examples for Two-Level Systems\label{Examples}}

\subsection{Optical Resonance with Spontaneous Emission}

We consider a two level atom in a classical resonant laser field. In 
the rotating-wave approximation the Hamiltonian for this system is
\begin{equation} 
        H=\hbar \begin{pmatrix} \frac{\Delta}{2} & \Omega
    e^{-i\phi} \\
    \Omega e^{i\phi} & -\frac{\Delta}{2} \end{pmatrix}
\end{equation}
The detuning $\Delta$, the coupling strength $\Omega$ and the phase
$\phi$ are properties of the laser.  To induce a geometric phase we
change the phase $\phi(t)$ of the laser field slowly in comparison to
$E/\hbar = (\Omega^2 + \frac{1}{4}\Delta^2)^\frac{1}{2}$, which is the
absolute value of the eigenenergies of the Hamiltonian divided by
$\hbar$.  The eigenvalue equation is
\begin{align} 
    H(t)\PLUS[(t)]
        &= E\PLUS[(t)] \\
        H(t)\MINUS[(t)] &= -E\MINUS[(t)]
\end{align}
with
\begin{align} 
    \PLUS[(t)]&=e^{-i\phi(t)\sin^2{\THETA}}\cos{\THETA}\EX + 
        e^{i\phi(t)\cos^2{\THETA}}\sin{\THETA}\GR \\
                \MINUS[(t)]&=-e^{-i\phi(t)\cos^2{\THETA}}\sin{\THETA}\EX +
        e^{i\phi(t)\sin^2{\THETA}}\cos{\THETA}\GR
\end{align}
and 
\begin{align} 
    \sin{\THETA}&=\sqrt{\frac{E-\half\hbar\Delta}{2E}} 
\label{theta1}\\
    \cos{\THETA}&=\sqrt{\frac{E+\half\hbar\Delta}{2E}} \label{theta2}
\end{align}
\EX and \GR denotes the excited state and the ground state of the two
level atom, respectively.  Note that \PLUS and \MINUS satisfy the
parallel transport condition as required in Section~\ref{MathMethods}.

Furthermore we want to include spontaneous emission as a source of 
dissipation. In the weak coupling limit the master equation is known 
to be 
\begin{equation} 
    \dot{\rho} (t) = \ih [H(t),\rho (t)] 
+\frac{1}{2}\mathcal{L}_\Gamma [\rho(t)]. 
%        \Gamma^\dagger -\half\Gamma^\dagger \Gamma \rho(t) - 
%      \half\rho(t)\Gamma^\dagger \Gamma 
\label{emission1}
\end{equation}
for
\begin{equation}
      \Gamma = \sqrt{\lambda}\:\sigma_- = 
    \sqrt{\lambda}\begin{pmatrix}0&0\\1&0\end{pmatrix}.
\end{equation}
Here $\lambda$ denotes the spontaneous emission rate.  The task here is
to solve \eqref{emission1} in the adiabatic and weak damping limit.  As
in Section \ref{MathMethods} we define the operator $A(t)$ by
$A(t)\PLUSM[(0)]=\PLUSM[(t)]$.  After the transformation of
\eqref{emission1} with $A^\dagger(t)$, the Hamiltonian is not diagonal.
Hence we carry out another transformation with an operator $B^\dagger$
that is defined by
\begin{align}
    B^\dagger |+(0)\rangle =& |e\rangle\notag\\
    B^\dagger |-(0)\rangle =& |g\rangle.
\end{align}
As $B^\dagger$ is time-independent, we obtain no term $B^\dagger\dot{B}$ in the
master equation.  We can carry out both transformations together with
the operator $C^\dagger(t)=B^\dagger A^\dagger(t)$, which turns out to
be
\begin{equation} 
        C(t)= 
        \begin{pmatrix} 
                e^{-i\phi(t)\sin^2{\THETA}}\cos{\THETA} &
                -e^{-i\phi(t)\cos^2{\THETA}}\sin{\THETA} \\
                e^{i\phi(t)\cos^2{\THETA}}\sin{\THETA} & 
                e^{i\phi(t)\sin^2{\THETA}}\cos{\THETA} 
        \end{pmatrix}.
\end{equation}
The master equation for $\rho^C(t)=C^\dagger(t)\rho(t)C(t)$ is, from
\eqref{meqn3} and \eqref{emission1},
\begin{equation} 
        \dot{\rho}^C (t) = \ih[H^C,\rho^C
        (t)]+\frac{1}{2}\mathcal{L}_{\Gamma^C (t)} [\rho^C(t)]
          \label{emission2}
\end{equation}
for
\begin{equation}
        H^C =  C^\dagger(t)H(t)C(t) = 
        \begin{pmatrix} 
                E&0\\0&-E 
        \end{pmatrix}
\end{equation}
and
\begin{align}
        \Gamma^C(t) = & C^\dagger(t)\Gamma C(t) \notag\\
        = & 
    \sqrt{\lambda}
    \begin{pmatrix}  
            \cos{\THETA}\sin{\THETA} & 
        \sin^2{\THETA}e^{-i\phi(t)\cos\theta} \\ 
        \cos^2{\THETA}e^{i\phi(t)\cos\theta} 
        & -\cos{\THETA}\sin{\THETA} 
    \end{pmatrix}.
\end{align}

We show in the Appendix that only the absolute value of the 
off-diagonal elements of the
dissipation operator $\Gamma^C(t)$ contributes in the solution of
\eqref{emission2}. Using only the absolute values of these elements 
does not simplify the calculation, but multiplying them with a time 
independent factor $e^{i\beta}$ and averaging over all 
$0 < \beta < 2\pi$ does.
Doing so, we first rewrite \eqref{emission2} 
\begin{align} 
    \dot{\rho}^C (t) =& \ih [H^C,\rho^C (t)] 
    + \frac{1}{4\pi}\int_0^{2\pi} \mathcal{L}_{\Gamma^C} 
[\rho^C(t)]
     \dd\beta . \label{emission2a}
\end{align}
In \eqref{emission2a} we substitute the Lindblad operator $\Gamma^C$ 
with the new Lindblad operators $\Gamma^C_\beta$ with 
\begin{equation} 
        \Gamma^C_\beta = \sqrt{\lambda}
    \begin{pmatrix}  
        \cos{\THETA}\sin{\THETA} & \sin^2{\THETA}e^{-i\beta} \\ 
        \cos^2{\THETA}e^{i\beta} & -\cos{\THETA}\sin{\THETA} 
     \end{pmatrix},\qquad 0\leq \beta < 2\pi
\end{equation}
as explained previously. Finally we get a new master equation, which 
is equivalent to \eqref{emission2} in the adiabatic and weak damping 
limit:

\begin{align} 
    \dot{\rho}^C (t) =& \ih [H^C,\rho^C (t)] 
    + \frac{1}{4\pi}\int_0^{2\pi} \mathcal{L}_{\Gamma^C_\beta (t)} 
[\rho^C(t)]
     \dd\beta \label{emission3}
\end{align}
\eqref{emission3} may seem to be more complicated than
\eqref{emission2}, but if we evaluate the term behind the integral in
\eqref{emission3} we realize that many contributions are cancelled out
by the integration.  If we write
\begin{equation}
    \rho^C(t)=\begin{pmatrix}a&b\\b^*&1-a\end{pmatrix}
\end{equation}
then \eqref{emission3} reduces to the two independent equations
\begin{align} 
    \dot{a}(t) &= 
        \lambda\Bigg[\sin^4{\THETA}
        -a\left(\sin^4{\THETA}+\cos^4{\THETA}\right)\Bigg]\\
        \dot{b}(t) &= -\frac{2i}{\hbar}Eb - \lambda 
        b\left(\sin^2{\THETA}\cos^2{\THETA}+\frac{1}{2} \right)
\end{align}
with the solutions
\begin{align} 
        a(t) =& \left(
        a(0)-\frac{\sin^4{\THETA}}{\sin^4{\THETA}+\cos^4{\THETA}}\right)
        e^{-\lambda t \left( \sin^4{\THETA}+\cos^4{\THETA} \right)} ,
                \nonumber       \\      &
        +\frac{\sin^4{\THETA}}{\sin^4{\THETA}+\cos^4{\THETA}} \\
        b(t) =& b(0) e^{-2i \frac{Et}{\hbar}} e^{-\lambda t 
        \left(\sin^2{\THETA}\cos^2{\THETA}+\half \right)}.
\end{align}
As the last step we need to evaluate
$\rho(t)=C(t)\rho^C(t)C^\dagger(t)$.  As the inversion provides an
operational quantity for inferring the geometric phase by measuring the
relative proportion of ground vs excited states, and because the terms
become rather long, we only write the inversion $w(t)$, 
which is
\begin{align} 
        w(t)=&\rho_{11}-\rho_{22}= (2a(t)-1)\cos\theta
                \nonumber       \\      &
        -2\sin\theta\text{Re}\left(
        b(t)e^{i\phi(t)\cos\theta}\right) \label{inversion}
\end{align}
To compare this result with that found by Ellinas et al \cite{Ell89}, we
set $\rho(0)=\half + p\:\sigma_3$ with $|p|\leq\half$ and substitute
for $\sin{\THETA}$ and $\cos{\THETA}$ from \eqref{theta1} and
\eqref{theta2}, respectively.  Furthermore we define
\begin{equation}
    K=\frac{2\Omega^2 +\Delta^2}{4\Omega^2 +\Delta^2}\text{ and }
    G=\frac{6\Omega^2+\Delta^2}{8\Omega^2+2\Delta^2}.
\end{equation}
If we furthermore consider the inversion at a time $T$ at the end of
the cyclic evolution we finally get for the inversion:
\begin{widetext}
\begin{align} 
        w(T) =  2p \Bigg( \left( \frac{\Delta\hbar}{2E}\right)^2
        \left(\frac{2Kp+1}{2Kp}e^{-K\lambda T}-\frac{1}{2Kp} \right)     
         + \cos{\left(\frac{2ET}{\hbar} 
        -2\pi\frac{\hbar\Delta}{2E}\right) } e^{-G\lambda T}
        \left( \frac{\Omega\hbar}{E}\right)^2 \Bigg)\label{Ell89Result}
\end{align}
which is the same as that derived by \cite{Ell89}.  The dynamic and
geometric phases are found in the cosine term in this expression: the
difference of the dynamic phases of the eigenstates of the
Hamiltonian is given by $2ET/\hbar$, and the difference of the
geometric phases of these eigenstates (for $\phi(T)=2\pi)$ given by
$2\pi \hbar \Delta/2E$.  This term is diminished by a damping factor
$\exp(-G\lambda T)$ which can influence the observability of the
geometric phase effect on the inversion.  The issues of time scales to
observe the effect of the geometric phase have been discussed in
\cite{Ell89}.  However, for the present, we wish to point out that the
result above has been derived here by use of a simple transformation
into a rotating frame.  This is to be contrasted with the much more
complicated approach of \cite{Ell89}, based on calculating the
eigenmatrices of the Liouvillian.

\subsection{Optical Resonance with Dephasing}

As in the previous subsection we treat a two level atom driven by a
resonant electromagnetic field.  This time we assume the damping is
due to dephasing that occurs as a consequence of phase changing 
collisions, which changes the relative phase between the excited state 
and the ground state of the atom (in contrast to strong collisions 
that change populations of eigenstates). Since the phase change can 
vary for each collision we have to consider a one dimensional 
manifold of dissipation operators
\begin{equation} 
        \Gamma_\alpha = \sqrt{\lambda(\alpha)} 
        \begin{pmatrix}
                1&0\\0&e^{i\alpha} 
        \end{pmatrix}
        ,\qquad -\pi < \alpha <\pi ,
\end{equation}
where $\lambda(\alpha)$ is the dephasing rate density.  Hence we get
the master equation
\begin{align} 
        \dot{\rho} (t) &= \ih [H(t),\rho (t)] + \INF \left(
        \Gamma_\alpha \rho(t) \Gamma^\dagger_\alpha -
        \lambda(\alpha)\rho(t) \right) \dd \alpha
        \label{dephasing1}
\end{align}
with the Hamiltonian $H(t)$ from the previous subsection with a slowly
changing phase $\phi(t)$ again.  Thus, we get the same parallel
transported eigenstates of the Hamiltonian and we can start by carrying
out the same transformation as in the previous subsection.  In the
adiabatic and weak damping limit we obtain the transformed master
equation
\begin{align} 
        \dot{\rho}^C (t) =& \ih [H^C,\rho^C (t)] 
        + \INF \left(
        \Gamma^C_\alpha(t) \rho^C(t) \Gamma^{C\dagger}_\alpha(t) -
        \lambda(\alpha)\rho^C(t) \right) \dd \alpha \label{dephasing2}
\end{align}
for
\begin{equation}       
        H^C =  C^\dagger(t)H(t)C(t) = 
        \begin{pmatrix} 
                E&0\\0&-E 
        \end{pmatrix}
\end{equation}
and
\begin{align}
    \Gamma^C_\alpha(t) &= C^\dagger(t)\Gamma_\alpha C(t)
        =
    \sqrt{\lambda(\alpha)} 
    \begin{pmatrix}
        \cos^2{\THETA}+e^{i\alpha}\sin^2{\THETA} & i
        \sin\theta\sin \frac{\alpha}{2} e^{i
        (-\phi(t)\cos\theta+\frac{\alpha}{2})} \\ i
        \sin\theta\sin \frac{\alpha}{2} e^{i
        (\phi(t)\cos\theta+\frac{\alpha}{2})} & \sin^2{\THETA} +
        e^{i\alpha}\cos^2\THETA 
        \end{pmatrix}.
\end{align}
Now everything is much the same as in the previous subsection.
Finally we find for the components $a$ and $b$
of the density operator $\rho^C(t)$ the decoupled differential
equations
\begin{align}
        \dot{a} &=-4fa\cos^2\THETA\sin^2\THETA +
        2f\cos^2\THETA\sin^2\THETA \\
        \dot{b} &=i\left(\frac{-2E}{\hbar} +
        gb\left(\sin^4\THETA - \cos^4\THETA \right)\right) -
        fb\left(\sin^4\THETA + \cos^4\THETA \right).
\end{align}
\end{widetext}
where 
\begin{equation}
        f=\int_{-\pi}^{+\pi} \lambda(\alpha)(1-\cos \alpha) \dd\alpha
\label{def:f}
\end{equation}
and
\begin{equation}
        g=\int_{-\pi}^{+\pi}
\lambda(\alpha)\sin\alpha \dd\alpha
\label{def:g}
\end{equation}
are properties of the model describing the damping collisions.  The
solutions of these equations are
\begin{align} 
        a(t) &= \left( a(0) - \half \right)
        e^{-4ft\cos^2\THETA\sin^2\THETA } +\half \label{a}, \\
        b(t) &= b(0) e^{i \left( \frac{-2E}{\hbar}+g\left(\sin^4 \THETA
        -\cos^4\THETA \right)\right)t} e^{-\left(\sin^4 \THETA
        +\cos^4\THETA \right) ft}.  \label{b}
\end{align}

To calculate the inversion~$w(t)$ we can take \eqref{inversion} and
substitute $a(t)$ and $b(t)$ with \eqref{a} and \eqref{b},
respectively.  Using \eqref{theta1} and \eqref{theta2} as well as the
previous definition of $K$ we finally find at time $t=T$:
\begin{align}
        w(T) =& 2p\Bigg( \left( \frac{\hbar\Delta}{2E}\right)^2
        e^{-\left(\frac{\Omega\hbar}{E}\right)^2fT} + \half
        \left(\frac{\Omega\hbar}{E}\right)^2 e^{-KfT}\nonumber\\
        &\times\cos \left( \left( g\frac{\hbar\Delta}{2E} -
        \frac{2E}{\hbar} \right)T + 2\pi\frac{\hbar\Delta}{2E}
        \right)\Bigg).  \label{dephasing3}
\end{align}
This result is similar to that found in the case of spontaneous
emission in \eqref{Ell89Result}.  There appears in the cosine term in
\eqref{dephasing3} the difference of the dynamic phases of the
eigenstates of the Hamiltonian, as given by $2ET/\hbar$, and the
difference of the geometric phases of these eigenstates (for
$\phi(T)=2\pi)$, given by $2\pi \hbar \Delta/2E$.  This term is also
diminished by a damping factor $\exp(-KfT)$ which influences the
observability of the geometric phase effect on the inversion.  Both
this term and an additional contribution of a shift in the Rabi
frequency by $g\frac{\hbar\Delta}{2E}$ arise through the presence of
dephasing, though the latter will only appear if the dephasing rate
density $\lambda(\alpha)$ is not symmetric.

\subsection{Spin in Magnetic Field with Dephasing}
As another example we consider the simple model of a spin-$\half$
particle in a magnetic field with constant field strength, which
demonstrates how to remove the dynamic phase in a standard model.  To
induce a geometric phase we change the direction of the magnetic field
slowly in comparison to $E/\hbar$.  As a source of decoherence we
consider dephasing which is defined by
\begin{align} 
    \Gamma_\alpha(t) \EX[(t)] &= \sqrt{\lambda(\alpha)}\EX[(t)] \\
    \Gamma_\alpha(t) \GR[(t)] &=
    \sqrt{\lambda(\alpha)}e^{i\alpha}\GR[(t)].
\end{align}
\GR[(t)] and \EX[(t)] are the parallel transported eigenstates of the
Hamiltonian with spin parallel and antiparallel to the magnetic field,
respectively.  Further $\lambda(\alpha)$ is the dephasing rate
density.  Note that dephasing does not change the energy of the
spin-system.  It is important to distinguish between this model and
the two level atom in the previous subsection.  Here the Lindblad
operators are defined in the basis of the time-dependent eigenstates
of the Hamiltonian whereas before the Lindblad operators have been
defined in the basis of the excited and the ground state of the two
level atom which are independent of the properties of the applied
laserfield and hence not the eigenstates of the Hamiltonian.  Such
dephasing operators could be realized by random fluctuations of the
field strength of the applied magnetic field.  Since the Hamiltonian
changes in time, the dephasing operators have to be time-dependent,
too.  As in Section \ref{MathMethods} we define the operator
\begin{align} 
    A(t) |e(0\rangle =&|e(t)\rangle\notag\\
    A(t)|g(0)\rangle =&|g(t)\rangle
\end{align}
and find, from \eqref{meqn3}, for $\rho^A = A^\dagger(t) \rho(t) A(t)$
\begin{align}
	\dot{\rho}^A(t) =& \ih \left[ 
    \begin{pmatrix} 
    	E&0\\ 0&-E
    \end{pmatrix},\rho^A(t) \right]\notag\\
        &+ \frac{1}{2}\INF \left[ 2
        \begin{pmatrix}
                1&0\\0&e^{i\alpha}
        \end{pmatrix} 
        \rho^A(t)
        \begin{pmatrix} 
                1&0\\0&e^{-i\alpha}
        \end{pmatrix} 
        - 2\rho^A(t) 
        \right]\nonumber \\ &\times\lambda(\alpha) \dd\alpha \notag \\
        =& \begin{pmatrix} 
                0 & \frac{-2i E}{\hbar} \rho^A_{12}(t) \\ 
                \frac{2i E}{\hbar} \rho^A_{21}(t) & 0
            \end{pmatrix} 
                \nonumber       \\      &
            + \INF 
        \begin{pmatrix} 
            0 & (e^{-i\alpha}-1) \rho^A_{12}(t) \\  
                (e^{i\alpha}-1)\rho^A_{21}(t) & 0
        \end{pmatrix} \nonumber \\ & \times \lambda(\alpha) 
        \dd\alpha. \label{spin2}
\end{align}
Here $\rho^A_{ij}(t)$ denotes the components of $\rho^A(t)$.  The
solution of \eqref{spin2} is 
\begin{align} 
        \rho^A_{11}(t) &=
        \rho^A_{11}(0) \\
        \rho^A_{12}(t) &= \rho^A_{12}(0) e^{-i\left( \frac{2 
E}{\hbar} +g \right)t} e^{ -ft } 
\end{align}
with $f$ and $g$ defined in Eqs.~(\ref{def:f}) and (\ref{def:g}),
respectively.  As the last step we need to calculate $\rho(t) =
A(t)\rho^A(t)A^\dagger(t)$.  If the evolution of the Hamiltonian is
cyclic we have $A(t)=$diag$(e^{i\varphi},e^{-i\varphi})$ where
$\varphi$ is the geometric phase for \EX. The geometric phase is half
of the solid angle enclosed by the path which \EX[(t)] drives on the
Bloch sphere \cite{Berry}.  This is equivalent to half of the solid
angle enclosed by the path determined by the direction of the magnetic
field.  Hence we finally get for the components of the density
operator after the Hamiltonian undergoes a closed loop
\begin{align} 
        \rho_{11}(T) &= \rho_{11}(0) \nn \\
        \rho_{12}(T) &= \rho_{12}(0) e^{-i\left( \frac{2 
        ET}{\hbar} +gT  - 2\varphi 
        \right)} e^{ -fT } \label{rho_12}
\end{align}

In the latter equation we can see a phase change due to the energy 
difference of the system, an additional phase change due to the 
dephasing and the geometric phase. Furthermore we see how the absolute
value of the off-diagonal element of the density operator decreases 
exponentially in time because of the dephasing.

Our task now is to remove the dynamic phase. We do a 
$\sigma_x$-transformation in our system and then in the time 
interval $[T,2T]$ we drive the direction of the magnetic field 
around the same loop as before but backwards: 
$\vec{B}(T+t)=\vec{B}(T-t)$. The components of the density operator 
$\rho'(T)$ after the $\sigma_x$-transformation are
\begin{align}
    \rho'_{11}(T) &= \rho_{22}(T) = 1-\rho_{11}(0) \nn\\
                \rho'_{12}(T) &= \rho_{21}(T) = \rho_{12}^*(0) 
e^{i\left(
                \frac{2 ET}{\hbar} +Tg - 2\varphi \right)}
                e^{-Tf}\label{forgot}
\end{align}
When we drive the magnetic field backwards, then the parallel 
transported eigenstates are $\EX[(T+t)]=\EX[(T-t)]$ and 
$\GR[(T+t)]=\GR[(T-t)]$.

Now we define the operator
\begin{align}
        A'(T+t)|e(T)\rangle &= |e(T+t)\rangle = |e(T-t)\rangle\notag\\
        A'(T+t)|g(T)\rangle &= |g(T+t)\rangle = |g(T-t)\rangle 
\label{second}
\end{align}
which parallel transports the eigenstates of the Hamiltonian
$H(T+t)$.  Again we transform the density operator
$\rho'^A(T+t) = A'^\dagger(T+t) \rho'(T+t) A'(T+t)$ and find for the
components of $\rho'^A(2T)$ \eqref{rho_12}
\begin{align}
        \rho'^A_{11}(2T) &=\rho'_{11}(T) = 1-\rho_{11}(0)  \nn\\
        \rho'^A_{12}(2T) &= \rho'_{12}(T) e^{-i\left( \frac{2
        E}{\hbar} +g\right)T} e^{ -Tf} = \rho_{12}(0)^* e^{-2i
        \varphi} e^{ -2Tf}
\end{align}
where in the last step \eqref{forgot} is used.  Now we need to find
$\rho'(2T) = A'(2T)\rho'^A(2T)A'^\dagger(2T)$.  From \eqref{second}, 
and
because now we drive the loop backwards and hence get the same
geometric phase up to a sign, it follows that
$A'(2T)=A^\dagger(T)=$diag$(e^{-i\varphi},e^{+i\varphi})$

After another $\sigma_x$-transformation we finally get the density 
operator
\begin{equation}
        \rho(2T)= \begin{pmatrix} \rho_{11}(0) & \rho_{12}(0) e^{4i
        \varphi} e^{ -2Tf } \\ \rho_{12}(0)^* e^{-4i \varphi} e^{
        -2Tf } & 1-\rho_{11}(0) \end{pmatrix}.
\end{equation}
Hence we see that not only the dynamic phase of the Hamiltonian is 
removed, but also the phase shift through dephasing. What stays is 
twice the difference of the geometric phases of the ground state and 
the excited state. This geometric effect appears in the off-diagonal 
components of the density operator and is damped out exponentially in 
time through dephasing.

\section{Generalizations\label{newdirections} }

\subsection{Non-Cyclic Evolution}
To consider a non-cyclic evolution we first outline Pati's analysis
\cite{Pati}.  If $H(T)=H(0)$, Pati compared the phase of the parallel
transported eigenstate of the Hamiltonian at time $t=T$, \N[(T)] with
the phase of the eigenstate at time $t=0$, \N[(0)].  If the Hamiltonian
does not undergo a closed loop, i.e. $H(T)\neq H(0)$, then \N[(T)] is
not \N[(0)] up to a geometric phase.  Comparing the phases of states
which differ not only by a phase is not straightforward.  Pati
introduced a reference section $\widetilde{\N[(t)]}$ which is supported
by eigenstates of the Hamiltonian $H(t)$.  The phase of
$\widetilde{\N[(t)]}$ is fixed by the requirement to make
$\widetilde{\N[(t)]}$ in phase with \N[(0)] as defined by means of the
work of \cite{Pan56}, i.e. $\NN[(0)] \widetilde{\N[(t)]} =0$.  Then
$\widetilde{\N[(T)]}$ and \N[(T)] differ only by a phase and this phase
is defined to be the generalization of the geometric phase to non-cylic
evolution.

We use this idea and generalize it to open systems. First one has to 
calculate the density operator $\rho^A(T)$ in the rotating axis 
representation as in Section~\ref{MathMethods}. Instead of transforming back to 
the original picture we transform to the picture given by the 
reference section introduced in \cite{Pati}. We define the operator 
$\tilde{A}(T)$ by
\begin{equation} 
    \widetilde{\N[(T)]} = \tilde{A}(T)\N[(0)]
\end{equation}
The density operator in this new picture is
\begin{align} 
    \rho^{\tilde{A}}(T) =& \tilde{A}^\dagger(T) A(T) \rho^A(T)
    A^\dagger(T) \tilde{A}(T) \nonumber \\ =& \text{diag}
    (e^{i\varphi_1}\! , \ldots ,e^{i\varphi_N})
    \rho^A(T)\text{diag}(e^{\! -i\varphi_1}\! , \ldots ,e^{\! 
-i\varphi_N})
\end{align}
and $\tilde{A}^\dagger(T) A(T) = \text{diag} (e^{i\varphi_1}, \ldots
,e^{i\varphi_N}) $ is the generalized holonomy transformation with
respect to the reference section $\widetilde{\N[(T)]}$.

\subsection{Non-Abelian Holonomies}

Again we consider the master equation \eqref{meqn1} with an
adiabatically changing Hamiltonian.  For simplicity we restrict to a
cyclic Hamiltonian $H(T)=H(0)$.

Until now we assumed the eigenenergies of the Hamiltonian to be
non-degenerate.  However, if the eigenvalues are degenerate we expect
to get non-Abelian holonomies \cite{WilczekZee, Anandan} as the
generalization of the geometric phase.  In this case we have the
eigenvalue equation
\begin{equation} 
    H(t) \N[_m(t)] = E_n(t) \N[_m(t)]
\end{equation}
in which $n=1,\cdots ,N$ ; $m=1,\cdots ,M_n$ and $M_n$ is the degree of
degeneracy of the subspace of the Hamiltonian with energy $E_n$.  The
$M_n$ are required to be constant in time, i.e.~we do not allow any
level crossings of the Hamiltonian $H(t)$.  The \N[_m(t)] are now
assumed to satisfy the modified parallel transport condition
\cite{Anandan}
\begin{equation}
        \NN[_m(t)]|\der \N[_{m'}(t)] =0 \qquad \forall\: m,m'=1, 
\cdots ,M_n.
        \label{new}
\end{equation}
Now we can define the operator $A(t)$ by
\begin{equation} 
        A(t) \N[_m(0)]=\N[_m(t)]
\end{equation}
and with its help we transform the master equation in the rotating
axis representation to remove the time dependence of the eigenspaces
of the Hamiltonian analogous to Section~\ref{MathMethods}.  Doing this
we get the new master equation \eqref{meqn2} for
$\rho^A(t)=A^\dagger(t)\rho(t)A(t)$ as before.  Again it can be shown
that the terms $\rho^A(t)A^\dagger(t)\dot{A}(t)$ and
$A^\dagger(t)\dot{A}(t)\rho^A(t)$ can be neglected if the \N[_m(t)]
satisfy the modified parallel transport condition \eqref{new} in the
adiabatic and weak damping limit.  This justifies the condition
\eqref{new}.  Since the proof for this is much the same as for
non-degenerate Hamiltonians as done in the Appendix, we do not carry
out the proof in this case.  Now we have to solve (16) which
represents the dynamics.  Finally we have to transform back to the
original picture:
\begin{equation}
  \rho(T)=A(T)\rho^A(T)A^\dagger(T).
\end{equation}
We obtain the non-Abelian holonomy $A(T)\in U(M_1)\otimes \cdots
\otimes U(M_N) $ simliar to the way obtained the geometric phase for
non-degenerate systems.

\section{Conclusions and Future Directions\label{conclusions}}

We have introduced the rotating axis transformation, in which
parallel transport of the eigenstates of the Hamiltonian plays an
important role, to study the geometric phase for an adiabatically
evolving multilevel system.  This transformation was shown to be
particularly useful in simplifying the calculation of open-system
evolution, described by a master equation of the
Lindblad form, as it allows an easy separation of dynamic and geometric
phases.  These advantages were illustrated by applying it to optical
resonance with spontaneous emission, where we obtain known results but
more easily.  The method was then used to quickly and easily study the
effects of the geometric phase in a number of new problems.  In one
application we show explicitly how to remove the dynamic phase.

Although, in our applications, we concentrated on Abelian holonomies 
for nondegenerate systems, the generalization to non-Abelian holonomies 
for degenerate Hamiltonians and to non-cyclic evolution is straight
forward. 

\acknowledgments

BCS acknowledges financial support from an Australian DEST IAP grant to
support participation in the European Fifth Framework project QUPRODIS
and from Alberta's informatics Circle of Research Excellence (iCORE) as
well as valuable discussions with S.~Ghose and K.-P.~Marzlin.

\appendix
\section{Neglecting $A^\dagger (t) \dot{A}(t)$ in the adiabatic 
approximation}

Here we prove that the terms containing $A^\dagger (t) \dot{A}(t)$ in
\eqref{meqn2} can be neglected in the adiabatic approximation.  The
proof will be analogous to the proof of the adiabatic theorem given in
\cite{Messiah}.  For simplicity we assume $H(t)$ in \eqref{meqn2} to 
be
diagonal which can always be achieved by a proper time independent
transformation and hence is no restriction.  We start by transforming
\eqref{meqn2} in the interaction picture.  We define
\begin{align} 
        \rho^H(t) &= e^{-i\int_0^t H^A(t) \dd t} \rho^A(t)
        e^{i\int_0^t H^A(t) \dd t} \nn\\
        \Gamma_\alpha^H(t) &= e^{-i\int_0^t H^A(t) \dd t}
        \Gamma_\alpha^A(t) e^{i\int_0^t H^A(t) \dd t} \nn\\
        (A^\dagger \dot{A})^H(t) &= e^{-i\int_0^t H^A(t)
        \dd t} A^\dagger (t) \dot{A}(t) e^{i\int_0^t H^A(t)
        \dd t} \nn
\end{align}
and get by use of \eqref{meqn2} the master equation in the 
interaction 
picture
\begin{equation} 
        \dot{\rho}^H = \rho^H(A^\dagger \dot{A})^H -
        (A^\dagger \dot{A})^H \rho^H 
        + \half\sum_{\alpha=1}^k \mathcal{L}_{\Gamma_\alpha^H}[\rho^H]
 \label{A1}
\end{equation}
The formal solution of this equation is
\begin{align} 
        \rho^H(t) =& \rho^H(0) 
        + \int_0^t \Bigg( \rho^H(A^\dagger \dot{A})^H -
        (A^\dagger \dot{A})^H \rho^H\Bigg. \notag\\
        &\Bigg.+ \half\sum_{\alpha=1}^k \mathcal{L}_{\Gamma^H_\alpha} 
[\rho^H] \Bigg) \!(s) \dd
        s.  \label{A2}
\end{align}
Within the integral, there are some contributions of the form of a
product of a slowly varying function and a fast oscillating function.
These contributions are known to become small when the frequency of
the oscillating function increases in comparison with the time
derivative of the slowly varying function.  To see this we make use of
the result, following \cite{Messiah},
\begin{align}
        \int_0^t f(s)e^{i\omega 
        s}\dd s & = \frac{1}{i\omega} \left( [f(s)e^{i\omega s}]^t_0 
        - \int_0^t f'(s)e^{i\omega s} \dd s \right) \nonumber \\ &
        \overset{\frac{\omega}{f'}\rightarrow 
\infty}{\longrightarrow}0 
\end{align}
To make use of this we write \eqref{A2} in components
\begin{align} 
        \rho^H_{ij}(t) = \rho^H_{ij}(0) &+ \int_0^t \Bigg( \rho^H_{ik}
        \,(A^\dagger \dot{A})^H_{kj} - (A^\dagger
        \dot{A})^H_{ik}\, \rho^H_{kj}\Bigg. \notag\\
        &\Bigg.+ \half\sum_{\alpha=1}^k
        2\Gamma^H_{\alpha ik}\,\rho^H_{kl}\,\Gamma^{H\dagger}_{\alpha
        lj} - \Gamma^{H\dagger}_{\alpha ik}\,\Gamma^H_{\alpha
        kl}\,\rho^H_{lj} 
        \nonumber \\ &- \rho^H_{ik}\,\Gamma^{H\dagger}_{\alpha
        kl}\,\Gamma^H_{\alpha lj} \Bigg) \!(s) \dd s.  \label{A3}
\end{align}
Here and later, summations are implied over all indices except of $i$
and $j$.  The components of $\dot{A}$ and $\Gamma$ are assumed
to be small in comparison with $\omega_{ij}= E_i-E_j$ (adiabaticity 
and
weak damping, respectively) and hence we see in \eqref{A3} that all
components of $\dot{\rho}^H$ are small in comparison with
$\omega_{ij}$.  The components of $(A^\dagger \dot{A})$ are
slowly varying and hence the off-diagonal components, $(A^\dagger
\dot{A})^H_{ij}$, $i\neq j$ are oscillating with frequency
$\omega_{ij}$ and can be neglected.  Because of the parallel transport
condition the diagonal elements $(A^\dagger \dot{A})^H_{nn}$ are
null as we can see:
\begin{align} 
        (A^\dagger \dot{A})^H_{nn} =& (A^\dagger \dot{A})_{nn}
        \nonumber       \\      
        =& \NN[(0)] | A^\dagger \dot{A} \N[(0)] = \NN[(t)] |\der
        \N[(t)] = 0 \nn
\end{align}
The last equality is true because we assumed the \N[(t)] to be 
parallel
transported.  So we have proved that we can neglect $A^\dagger
\dot{A}$ in \eqref{meqn2}.  

Furthermore we can rewrite \eqref{A3} as
\begin{align} 
        \rho^H_{ij}(t) = &\rho^H_{ij}(0) + \int_0^t \Big(
        \half\sum_{\alpha=1}^k 2\Gamma^H_{\alpha
        ik}\,\rho^H_{kl}\,\Gamma^{H *}_{\alpha jl} 
                        \nonumber       \\      &
        - \Gamma^{H *}_{\alpha
        ki}\,\Gamma^H_{\alpha kl}\,\rho^H_{lj} - 
\rho^H_{ik}\,\Gamma^{H
        *}_{\alpha lk}\,\Gamma^H_{\alpha lj} \Big) \!(s) \dd s.
\end{align}
The star denotes complex conjugation.  The functions
$\Gamma^H_{ik}\Gamma^{H*}_{jl}$ are oscillating with frequency
$\omega_{ik}-\omega_{jl}$.  Hence we can achieve a significant
simplification if the differences of all eigenfrequences,
$\omega_{ik}-\omega_{jl}$ are not vanishing (are big in comparison 
with
$\dot{H}$ and $\Gamma_\alpha$) which is always the case if we
consider a two-level system.  Then the condition
$\omega_{ik}-\omega_{jl}=0$ implies $i=k \,, j=l$ or $i=j \,, k=l$ and
hence only corresponding parts will contribute to the integral:
\begin{align} 
        \rho^H_{ij}(t) =& \rho^H_{ij}(0) + \int_0^t \!\Bigg(
        \!\half\sum_{\alpha=1}^k 2 \delta_{ij} \Gamma^H_{\alpha
        ik}\,\rho^H_{kk}\,\Gamma^{H *}_{\alpha ik} 
        \nonumber \\ & - 2 \delta_{ij}
        \Gamma^H_{\alpha ii}\,\rho^H_{ii}\,\Gamma^{H *}_{\alpha
        ii} + 2
        \Gamma^H_{\alpha ii}\,\rho^H_{ij}\,\Gamma^{H *}_{\alpha jj} 
        \nonumber \\ &-
        \Gamma^{H *}_{\alpha ki}\,\Gamma^H_{\alpha ki}\,\rho^H_{ij} -
        \rho^H_{ij}\,\Gamma^{H *}_{\alpha lj}\,\Gamma^H_{\alpha lj}
        \!\!\Bigg) \!(s) \dd s.
\end{align}
Here we see that only the absolute value of the off-diagonal elements 
of $\Gamma^H_\alpha$ and hence  $\Gamma^A_\alpha$ contribute in 
\eqref{meqn2}.

We can further note that if we set $i=j$, we find that the diagonal 
elements of the density operator are coupled only to diagonal 
elements, whereas for $i\ne j$, we find that $\rho_{ij}^{H}$ is 
coupled only to itself.

\end{document}